\documentclass[aps,showpacs,pra,reprint,groupedaddress,longbibliography,amsmath,amssymb,superscriptaddress,intlimits]{revtex4-2}

\usepackage{amssymb,amsmath}
\usepackage{bm}
\usepackage[hidelinks]{hyperref}
\usepackage{graphicx}
\usepackage{txfonts}
\usepackage{xcolor}
\usepackage{mathtools}
\usepackage{makecell,tabularx}
\usepackage{siunitx}
\usepackage[utf8]{inputenc} 

\newcommand{\ee}{\mathrm{e}}
\newcommand{\ii}{\mathrm{i}}
\newcommand{\ff}{\mathrm{f}}

\setcellgapes{3pt}

\begin{document}

\title{Adiabatic Quantum Zeno Dynamics of Bosonic Atom Pairs with Large Inelastic Losses}

\author{
Manel Bosch Aguilera$^{1}$, 
Alexis Ghermaoui$^{1}$, 
R\'{e}my Vatr\'e$^{1}$, 
Rapha\"{e}l Bouganne$^{1}$, 
J\'{e}r\^{o}me Beugnon$^{1}$, 
Fabrice Gerbier$^{1}$
\email[]{fabrice.gerbier@lkb.ens.fr}
}

\affiliation{$^1$Laboratoire Kastler Brossel,  Coll\`ege de France, CNRS, ENS-PSL University, Sorbonne Universit\'e, 11 Place Marcelin Berthelot, 75005 Paris, France}

\date{\today}

\begin{abstract}
 We report on experiments exploring the non-Hermitian dynamics of pairs of two-level atoms tightly confined in an optical lattice and driven by a near-resonant laser. Although spontaneous emission is negligible for the long-lived excited state, two-body dissipation arises from strong inelastic collisions between two atoms in the excited state. We demonstrate quasi-adiabatic control of the internal state of the pairs in the quantum Zeno regime where inelastic losses dominate.  Preparing each atom pair in the longest-lived eigenstate of the non-Hermitian Hamiltonian describing the dissipative dynamics, we measure a lifetime enhanced by more than two orders of magnitude in comparison to the bare two-body lifetime. The measured enhancement factor is in quantitative agreement with the expected lifetime of the prepared ``non-Hermitian dressed state''.  
\end{abstract}

\maketitle

\section{Introduction}
In recent years, the theory of non-Hermitian systems has emerged as a common framework to describe classical waves undergoing energy or information loss. Experiments in optical, mechanical, biological or electrical systems\,\cite{Ashida20} have demonstrated many phenomena unique to non-Hermitian systems, such as unidirectional transport, enhanced sensing capability or characteristic topological properties in the vicinity of the so-called \textit{exceptional points} where two (or more) of the eigenmodes coalesce\,\cite{dembowski01,doppler16,xu16}. 

Non-Hermitian Hamiltonians\,\cite{moiseyev11,brody2013a} have also been used for a long time to model dissipative quantum systems, starting with early studies of radioactive decay\,\cite{gamow28,siegert39}. Dissipation in quantum systems often originates from their weak coupling to a larger environment, described by a Lindblad master equation\,\cite{Cohen-Tannoudji1992},
\begin{align}\label{eq:Meq}
\frac{d \hat{\rho}}{dt} & = \frac{1}{\ii \hbar} \Big[ \hat{H}_{\mathrm{eff}}, \hat{\rho} \Big] + \sum_\alpha \,  \Gamma_\alpha\, \hat{L}_\alpha \,\hat{\rho}\, \hat{L}_\alpha^\dagger.
\end{align}
Here $\hat{\rho}$ is the density matrix, $ \hat{L}_\alpha$ is the so-called jump operator for the dissipative channel $\alpha$, and $\hat{H}_{\mathrm{eff}}=\hat{H}-\ii \sum_\alpha \frac{\hbar \Gamma_\alpha}{2} \hat{L}_\alpha^\dagger \hat{L}_\alpha$ is the effective Hamiltonian, with the Hermitian part $\hat{H}$ describing the isolated system. The jump operator $ \hat{L}_\alpha$ describes environment-induced transitions in the dissipative channel $\alpha$ at an average rate $\Gamma_\alpha$. The quantum trajectory approach\,\cite{Dalibard92,Dum92,Daley14} interprets the non-Hermitian operator $ \hat{H}_{\mathrm{eff}}$ as an effective Hamiltonian generating a continuous dissipative evolution interrupted by stochastic quantum jumps described by the $\hat{L}_\alpha$'s. Thus, in general, the non-Hermitian Hamiltonian captures only part of the dissipative dynamics. 

Master equations such as Eq.\,(\ref{eq:Meq}) have a long history and many applications in low-energy physics, in particular in atomic physics and quantum optics\,\cite{Cohen-Tannoudji1992,Muller12,Daley14}. More recently, master equations and non-Hermitian Hamiltonians have been used to study many-body systems out of equilibrium. These studies have been largely driven by experimental progress, for instance with ultracold atoms subject to strong one-, two- or three-body losses\,\cite{Syassen08,GarciaRipoll09,Duerr09,barontini13,Yan13,Zhu14,Labouvie16,Patil15,Luschen17,Tomita17,Sponselee18,Schemmer2018a,Dogra2019a,Li19,Tomita19,Mark20}. The dissipative part of the Lindblad master equation can be interpreted as describing generalized measurements performed on the system\,\cite{gagen93,cirac94,javanainen13}. For strong losses, one enters a regime where the quantum Zeno effect\,\cite{Itano90,Fischer01,Streed06} -- the freezing of the time evolution of a quantum system subject to frequent measurements --becomes relevant. This effect prevents transitions to the highly dissipative states, thereby effectively suppressing dissipation and dramatically slowing down the decay\,\cite{Syassen08,GarciaRipoll09,Duerr09,barontini13,Yan13,Zhu14}.   
Atomic systems with strong two-body losses are plentiful, for instance two-electron atoms (as in our work), but also atoms in highly excited Rydberg states, atoms close to a Feshbach resonance, or spinful dipolar atoms ... These systems also have interesting properties that have led to many original proposals to engineer correlated quantum states or more generally study quantum many-body physics\,\cite{gorshkov09,gorshkov10,foss-feig10,gerbier10,olmos13,cazalilla14}. 
 
In this article, we investigate the dissipative properties of pairs of two-level atoms trapped at isolated nodes of an optical lattice and driven by a near-resonant laser\,\cite{Bouganne17,Franchi17}. The two atoms interact elastically in a state-dependent manner, but also undergo strong inelastic losses when both atoms are excited. In a quantum trajectory description, quantum jumps project the quantum state at each lattice site to the vacuum state, which does not contribute to any observable. As a result, expectation values are completely determined by the evolution under the non-Hermitian Hamiltonian.
%
We demonstrate quasi-adiabatic\,\cite{nenciu92,Uzdin11,Ibanez14,milburn15,nasari2022a,Melanathuru2022a,singhal2022a} control of the quantum state of the pairs in the quantum Zeno regime of strong dissipation. We prepare each atom pair in the longest lived ``dressed'' eigenstate of the effective non-Hermitian Hamiltonian, and observe a lifetime longer than the bare two-body lifetime by at least two orders of magnitude. Preparing (quasi-)adiabatically many-body systems in loss-protected states is of paramount importance to realize experimentally the proposals discussed above. The experiments presented here are a first step towards this goal.



%
%
%
%
%

\section{Dynamics of laser-driven, interacting atom pairs and non-Hermitian Hamiltonian}
\label{sec:theory}

\begin{figure*}[ht!!!!!]
\centering
\includegraphics[width=2\columnwidth]{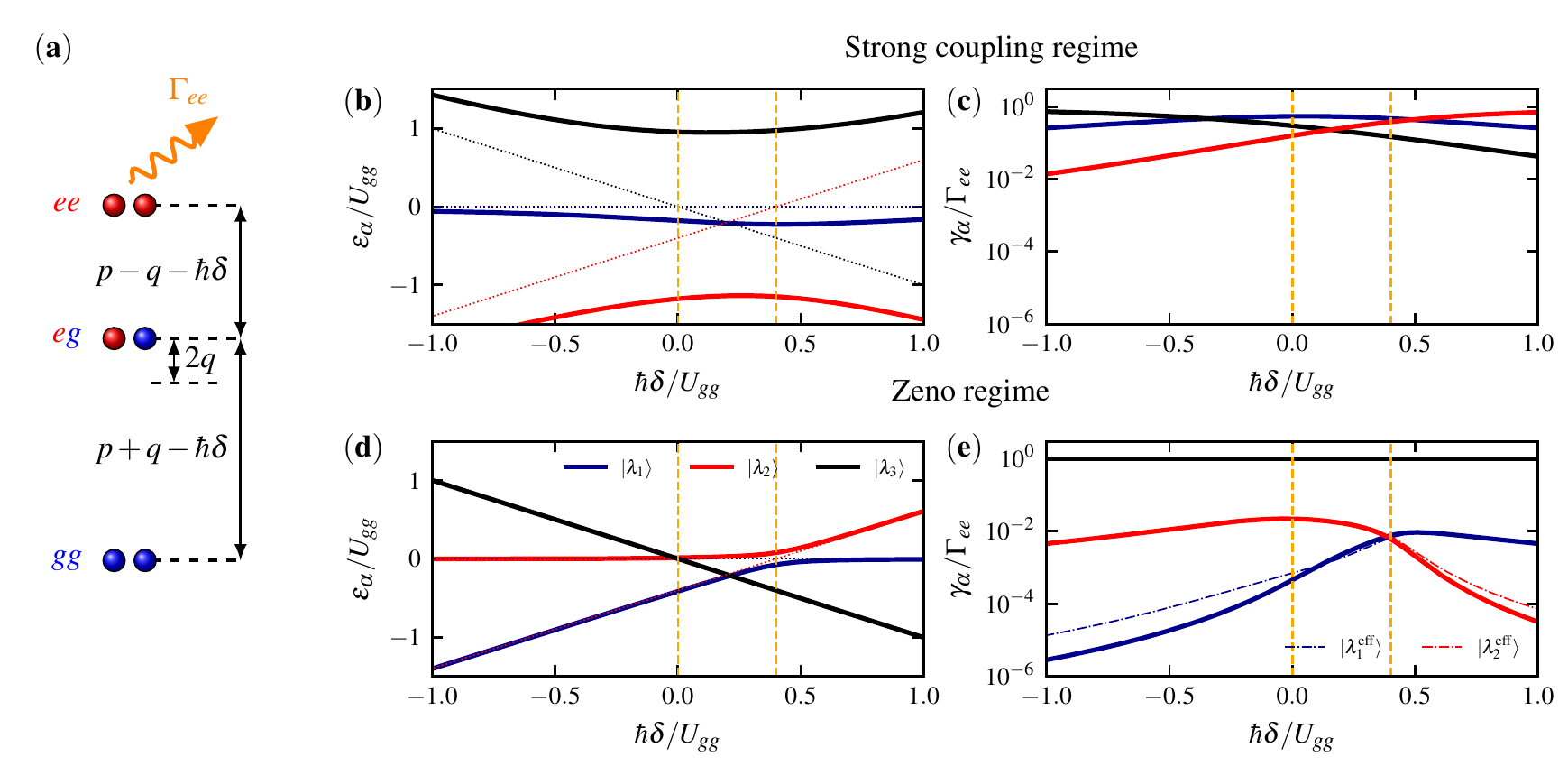}
\caption{(a): Sketch of the energy levels of a pair of bosonic atoms with internal states $g,e$. The parameters $p =(U_{ee}-U_{gg})/2$ and $q=U_{eg}-(U_{gg}+U_{ee})/2$ characterize the shifts of the transition frequencies due to state-dependent interactions $U_{\alpha\beta}$ ($\alpha,\beta=\{g,e\}$), $\delta$ is the laser detuning from the atomic resonance, and $\Gamma_{ee}$ is a two-body loss rate for two atoms in the excited state $e$. (b-c): Real and imaginary parts of the eigenvalues of the effective Hamiltonian in the regime of weak dissipation with $\Omega/\Gamma_{ee} = 1$.  (d-e): Same in the quantum Zeno regime $\Omega/\Gamma_{ee} = 0.1$. The thin dotted lines in (b,d) indicate the uncoupled energy levels $gg, eg, ee$. The vertical dashed lines in (b-e) mark the position of the one-photon resonances. The dash-dotted lines in (e) indicate the decay rates calculated from the effective Hamiltonian in the $\{gg,eg \}$ subspace [Eq.\,(\ref{eq:lambda12_Heff})]. We used the experimental values from Table\,\,\ref{tab1}. }\label{fig1}
\end{figure*}

\subsection{The non-Hermitian Hamiltonian}


We consider a pair of bosons with two internal levels confined to the motional ground state of a tight trap. A near-resonant laser field of
 frequency  $\omega_{\mathrm L}$  couples the two internal states $\vert g \rangle$ and $\vert e\rangle$. Neglecting motional excitations, the internal state is the only dynamical degree of freedom. The two-particle Hilbert space has only three exchange-symmetric states, that we label $\vert gg\rangle,\vert eg\rangle$ and $\vert ee\rangle$ to highlight the occupation numbers of the internal states. 
The dynamics is captured by an \textit{effective} non-Hermitian Hamiltonian  (see Supplemental Material\,\cite{supmat}),
\begin{align}\label{eq:Heff}
\hat{H}_{\mathrm{eff}}&=\hat{H}_0 \, + \, \hat{W} \, -\ii \, \frac{\hbar\Gamma_{ee}}{2}\,\vert ee \rangle \langle ee \vert.
\end{align}
The first two terms in the right hand side (rhs) describe coherent internal dynamics, with 
\begin{align}\label{eq:H0}
\hat{H}_0  &= \big(p-\hbar\delta) \hat{S}_z - q \hat{S}_z^2,
\hspace{0.3cm}
\hat{W} = 
\hbar\Omega\hat{S}_x.
\end{align} 
Here the coupling strength is given by the Rabi frequency $\Omega$, the detuning from the one-atom transition frequency $\omega_{0}$ is $\delta=\omega_{\mathrm L}-\omega_{0}$, and
\begin{align}
\hat{S}_z=
\begin{pmatrix} 
-1 & 0 & 0\\
0 & 0 & 0\\
0 & 0 & 1
\end{pmatrix},
\hspace{0.3cm}
\hat{S}_x=\frac{1}{\sqrt{2}}
\begin{pmatrix} 
0 & 1 & 0\\
1 & 0 & 1\\
0 & 1 & 0
\end{pmatrix},
\end{align} 
are the standard spin-1 matrices. The uncoupled Hamiltonian $\hat{H}_0 $ includes the internal energies and the interaction energies of the pair for each spin configuration $\vert ij \rangle$,
\begin{align}\label{eq:Uij}
U_{ij} &=\frac{4\pi \hbar^2 a_{ij} }{M}\int \vert w(\mathbf{r}) \vert^4 \, d^3\mathbf{r} ,
\end{align}
with $M$ the mass of an atom, $a_{ij}$ the scattering length describing the particular interaction process, and $w$ the orbital wavefunction. The relevant shifts of the transition frequencies are parametrized by   $p =(U_{ee}-U_{gg})/2$ and $q=U_{eg}-(U_{gg}+U_{ee})/2$. Relevant values of the various parameters for our experiment are indicated in Table\,\ref{tab1}. 

The last term in the rhs of Eq.\,(\ref{eq:Heff}) proportional to the projector $\vert ee \rangle \langle ee \vert$ describes inelastic losses. We neglect two-body loss processes for the $gg$ and $eg$ states, in accordance with experimental measurements\,\cite{Bouganne17,Franchi17}. Physically, the losses originate from ``principal quantum number changing'' collisions\,\cite{kelly88,traverso09} (also called ``energy-pooling'' collisions), concisely summarized by the reaction $e +e \to g+e' $ with $e'$ another electronic excited state which eventually radiatively decays to $g$. The internal energy released in this strongly inelastic process makes both atoms escape from the trap.  The two-body loss rate
\begin{align}
\Gamma_{ee} &=\beta_{ee} \, \int \vert w(\mathbf{r}) \vert^4 \, d^3\mathbf{r},
\end{align}
with $\beta_{ee}$ an atom-dependent rate constant, is determined by the same overlap integral as the elastic interaction in Eq.\,(\ref{eq:Uij}). As a result, the ratio  $\hbar \Gamma_{ee}/U_{gg}=M \beta_{ee}/(4\pi \hbar a_{gg})$ is independent of the lattice details and of order unity (see Table\,\ref{tab1}).  

\subsection{Properties of the non-Hermitian Hamitlonian}
%

We consider how the complex eigenspectrum $\lambda_n=\varepsilon_n - \ii \hbar\gamma_n/2$ (with $\varepsilon_n $ and $\gamma_n$ real numbers) $\hat{H}_{\mathrm{eff}}$ changes as a function of the detuning $\delta$. We call the eigenstates $\vert \lambda_n\rangle$ ``non-Hermitian dressed states'' in analogy with their Hermitian counterparts\,\cite{supmat}. Note that the imaginary part (``decay rate'') of the non-Hermitian spectrum is related to the population of the bare state $ee$,
\begin{align}\label{eq:gamma_i}
 \gamma_n  = - \frac{2}{\hbar}  \mathrm{Im}(\lambda_n) =   \Gamma_{ee}  \Big\vert\Big\vert \langle ee \vert\lambda_n \rangle \Big\vert\Big\vert^2.
\end{align}

To set the stage, we first discuss the behaviour of the coherent part $\hat{H}_0 \, + \, \hat{W}$. The spectrum of the Hamiltonian $\hat{H}_0$ (thin dotted lines in  Fig.\,\ref{fig1}b,d) gives rise to three level crossings. Two of these (marked by the thin vertical dashed lines in Fig.\,\ref{fig1}b-e) correspond to one-photon transitions and occur when $\delta=(p+q)/\hbar$ ($gg-eg$ crossing) and when $\delta=(p-q)/\hbar$ ($eg-ee$ crossing). The third crossing corresponds to a two-photon transition and occurs when $\delta=p/\hbar$ ($gg-ee$ crossing). A non-zero laser coupling $\hat{W}$ turns the level crossings to avoided crossings.  

Turning to the spectrum of the complete non-Hermitian $\hat{H}_{\mathrm{eff}}$, we find that two regimes emerge either at strong ($\Omega \gtrsim \Gamma_{ee}$) or weak ($\Omega \ll \Gamma_{ee}$) coherent driving, respectively. For strong driving, the eigenstates are qualitatively similar to the situation without losses. The real parts $\epsilon_i$ display avoided crossings as for the non-dissipative version of the problem (Fig.\,\ref{fig1}b). Near resonance ($\vert \delta \vert \lesssim \Omega$), the dressed states are superpositions of all three bare states with comparable weigths, and all decay rates $\gamma_i$ are comparable to the bare loss rate $\Gamma_{ee}$ (Fig.\,\ref{fig1}c). 

For weak driving (or, equivalently, strong losses), a qualitatively different situation emerges. Only the $gg-eg$ avoided crossing survives. The other two disappear and the real parts $\epsilon_i$ cross as in the absence of laser coupling (Fig.\,\ref{fig1}d). Moreover, the non-Hermitian dressed states no longer mix near resonance as for strong driving (Fig.\,\ref{fig1}e). Instead, one notices that the states $\{  \lambda_1, \lambda_2 \}$ form a ``lossless'' subspace with a decay rate orders of magnitude smaller than the bare loss rate $\Gamma_{ee}$. As the decay rate is varied, this subspace remains well isolated from the lossy $ \lambda_3$ state with a decay rate $\gamma_3 \approx \Gamma_{ee}$ for all detunings. 

\subsection{Effective Hamiltonian in the lossless subspace}
%

In the quantum Zeno regime, the lossy state remains isolated in the sense that the complex eigenvalue $\lambda_3$ never approaches $\lambda_{1/2}$ because of the vastly different imaginary parts. This is reminiscent of a common situation in quantum mechanics, where a low-energy subspace remains well isolated in energy from the rest of the Hilbert space. The dynamics in the low-energy subspace can be captured by an \textit{effective Hamiltonian} 
given by the expansion\,\cite{Cohen-Tannoudji1992},
\begin{align}
\hat{P} \, \hat{H}_{\mathrm{eff}}' \, \hat{P} - \hat{P} \, \hat{H} \, \hat{P}& = \hat{P} \hat{H} \hat{Q} \, \frac{1}{\hat{H}} \, \hat{Q} \hat{H} \hat{P} + \cdots,
\end{align}
where $\hat{Q}$ and $\hat{P}=\mathbf{\hat{1}}-\hat{Q}$  are projectors on the high- and low-energy subspaces.

The formalism of the effective Hamiltonian is easily adapted to the non-Hermitian case, with \textbf{$\hat{Q}=\vert ee \rangle \langle ee \vert$} and $\hat{P}=\mathbf{\hat{1}}-\hat{Q}$ the projectors on the lossy and lossless subspaces, respectively, and with $\hat{P} \hat{W} \hat{Q}=(\hbar\Omega/\sqrt{2}) \times  \vert eg \rangle \langle ee \vert $. The effective Hamiltonian in the lossless subspace spanned by $\{ \vert gg \rangle, \vert eg \rangle \}$ is given by 
\begin{align}
\hat{P} \, \hat{H}_{\mathrm{eff}}' \, \hat{P} & = \frac{\hbar}{2}\begin{pmatrix} 
\delta'  & \sqrt{2}\Omega\\
 \sqrt{2}\Omega &-\delta'- \ii \Gamma_{\rm eff} \end{pmatrix}.
\end{align}
with $ \delta'=\delta-(p+q-\Delta q)/\hbar$.  The $eg$ level acquires an additional level shift $\Delta q$ and an effective decay rate $\Gamma_{\mathrm{eff}}$,
\begin{align}
\label{eq:gammaeff}
\Delta q + \ii \frac{\Gamma_{\mathrm{eff}}}{2} &=   \frac{\Omega^2}{\Gamma_{ee}} \, \frac{x+ \ii}{ 1+x^2},
\end{align}
with $x=2(p-q-\hbar\delta)/(\hbar\Gamma_{ee})$. The small parameter of the expansion is $\Omega/\Gamma_{ee} \ll 1$ in the quantum Zeno regime, so that $\Gamma_{\mathrm{eff}}, \, \vert q'-q\vert/\hbar \ll \Omega$. For the experimental parameters $\Omega/(2\pi) \approx 150\,$Hz and $\Gamma_{ee}/(2\pi) \approx 1.5\,$kHz, we have for instance $\Gamma_{\mathrm{eff}} /(2\pi) \approx 60\,$s$^{-1}$ on resonance ($x=0$). For  $\Omega \ll \Gamma_{ee} $, the eigenvalues of the $2 \times 2$ effective Hamiltonian are approximately
\begin{align}\label{eq:lambda12_Heff}
\lambda_{1/2} \approx \pm \frac{\hbar \Omega_{2\times2}}{2}  -\ii \frac{\hbar\Gamma_{\mathrm{eff}}}{4} \left(1 \mp \frac{\delta'}{\Omega_{2\times2}}\right).
\end{align}
with $\Omega_{2\times2}=\sqrt{\delta'^2+2\Omega^2} $. The imaginary parts of the two eigenvalues are shown in Fig.\,\ref{fig1}e. 

Far off resonance ($\vert x \vert \approx 2\vert\delta\vert/\Gamma_{ee} \gg 1$), the effective decay rate $\Gamma_{\mathrm{eff}} $ behaves as $\Gamma_{\mathrm{eff}} \propto  \Gamma_{ee} (\Omega/\delta)^2$, which is simply the bare decay rate $\Gamma_{ee}$ times the occupation probability $\propto (\Omega/\delta)^2$ of the $ee$ state in the large detuning limit. The decay of the slowest decaying state is even smaller and scales as $\propto (\Omega/\delta)^4$. The scaling reflects that for large detunings, the slowest decaying state couples to  $ee$  by a two-photon transition. 

As the detuning is brought closer to the $\delta'=0$ resonance, the parameter $\vert x \vert $ eventually becomes small compared to one, and the decay rate approaches $\Gamma_{\mathrm{eff}}/2 \propto  \Omega^2/ \Gamma_{ee} \ll \Gamma_{ee}$ for both eigenstates. Moreover, the larger $\Gamma_{ee}$ the smaller the saturation value, a behavior typical of the quantum Zeno effect. Note that a large bare decay rate is essential. Indeed, in the opposite limiting case where $\vert x \vert \gg 1$ near resonance (or equivalently, $ 2\vert q\vert \gg \hbar \Gamma_{ee}$), one finds instead a perturbative scaling law $\Gamma_{\mathrm{eff}} \propto  \Gamma_{ee}$.

\begin{table}
\makegapedcells

\setlength\tabcolsep{6pt}

    \begin{tabularx}{\linewidth}{cccccc}
   \Xhline{0.8pt}
$U_{gg}/h$ & $U_{eg}/U_{gg}$ & $U_{ee}/U_{gg}$ & $\hbar\Gamma_{ee}/U_{gg}$& $p/U_{gg}$ & $q/U_{gg}$\\
    \Xhline{0.3pt}
   \hline
$1.40\,$kHz & 0.905 & 1.21 & 1.02 & 0.21 & 0.20\\
    \Xhline{1pt}
\end{tabularx}
\caption{Summary of collisional properties. We calibrate experimentally the interaction strength $U_{gg}$ for ground state atoms as in \cite{Bouganne17}, while $U_{eg}$, $U_{ee}$ and $\Gamma_{ee}$ are deduced from it using the measured scattering lengths\,\cite{Franchi17} and inelastic rate constant\,\cite{Bouganne17}. The quantities $p=U_{ee}-U_{gg}$ and $q=(U_{ee}+U_{gg})/2-U_{eg}$ determine the shifts of the transition frequencies (see Fig.\,\ref{fig1}).}
\label{tab1}
\end{table}

\section{Description of the experiments}

\subsection{Experimental system}

\begin{figure}[ht!!!!!!]
\centering
\includegraphics[width=\columnwidth]{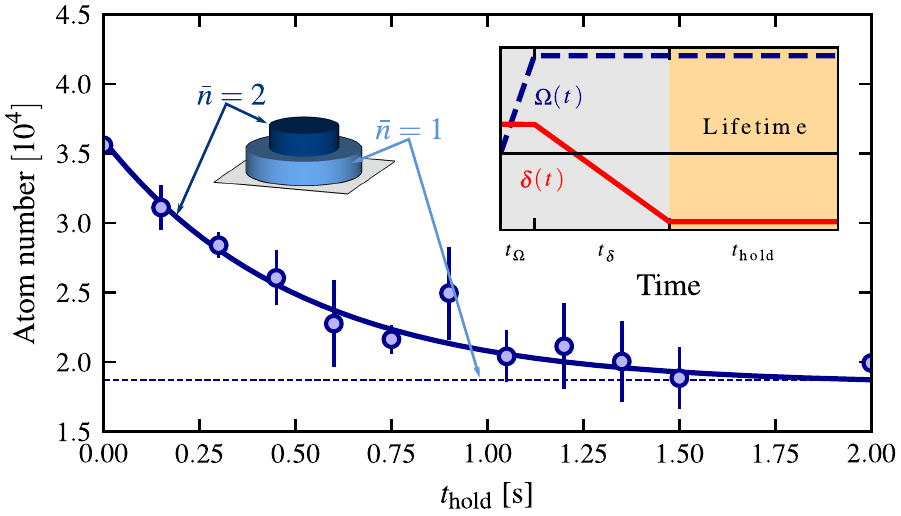}
\caption{Lifetime measurement for a final detuning $\delta_\ff /(2\pi) = 650\,$Hz and nominal Rabi frequency $\Omega /(2\pi) = 150\,$Hz. The figure shows the total atom number (irrespective of the internal state) versus hold time. The decay is attributed to two-body losses in doubly-occupied sites initially occupying the center of the trap. The presence of an outer shell of singly-occupied sites explains the non-zero asymptote. The inset shows a sketch of the temporal profile of the ramp. }
\label{fig2}
\end{figure}

The experiment is performed with ultracold $^{174}$Yb atoms trapped at isolated sites of a three-dimensional optical lattice. 
The optical lattice operates at the magic wavelength $\lambda_{\mathrm{m}} \approx 759.4\,$nm, where the lattice potential is independent of the internal state $g$ or $e$. The lattice depth is chosen deep in the Mott insulator regime with depth $V_{x,y,z}=25,25,27\,E_r$, where $E_r=h \times 1.98\,$kHz is the lattice recoil energy. We assume that band occupation beyond the fundamental Bloch band (including interband transitions induced by the coupling laser) and tunneling within that band are negligible on experimental time scales.  In this regime, the lattice sites can be treated as an ensemble of tight traps isolated from each other. Due to an auxiliary harmonic trap\,\cite{Bouganne17}, the system forms a core of doubly-occupied sites near the trap center, surrounded by a shell of singly-occupied (or empty) sites (see sketch in Fig.\,\ref{fig2}). Our focus is on the dynamics of atom pairs populating the central ``Mott core'', as considered in the previous Section\,\ref{sec:theory}. 


\subsection{Landau-Zener ramps}

The atoms initially occupy the electronic ground state $g \equiv$$ ^1\textrm{S}_0$, coupled to the excited state $e \equiv$$ ^3\textrm{P}_0$ by a narrow-linewidth ``clock'' laser with a wavelength around 578\,nm. We prepare the system to the desired final state using ``Landau-Zener ramps'' where the detuning changes linearly with time (inset of Fig.\,\ref{fig2}). In details, the laser is first turned on with a negligible intensity at an initial detuning $\delta_\ii$ far from any resonance. A first intensity ramp  of duration $t_\Omega$ brings the Rabi frequency to its nominal value $\Omega=2\pi \times 150\,$Hz, keeping $ \delta = \delta_\ii$ fixed. A second ramp of duration $t_\delta$ brings the detuning to the final desired value $\delta_\ii\to\delta_\ff$, keeping $\Omega$ fixed at the nominal value. For all reported experiments, we perform the frequency ramp at a constant speed $\dot{\delta}\approx 2\pi \times 11.1\,$Hz/ms, so that $t_\delta=(\delta_\ff-\delta_\ii)/\dot{\delta}$, and chose $t_\Omega=T_R/10$, for a total ramp time $T_R = t_\Omega+t_\delta$.


\subsection{Detection}\label{sec:detection}

We measure either the total atom number or the populations in the bare state $g$ or $e$ using absorption imaging after a time of flight\,\cite{supmat}. The $e$ population is measured up to a global repumping efficiency $\eta_{\rm rp}\approx 0.8$\,\cite{Bouganne17}. The result of a given measurement can be expressed in terms of the occupation probability $P_\alpha$ of the bare atomic states, weighted by the initial proportions of singly- or doubly-occupied sites. For instance, the population of state $g$ is given by  
\begin{align}\label{eq:Ng}
N_g & = n_1 \, P_{g}^{\overline{n}=1} \, + \, n_2 \, \left(  P_{eg} +  2 P_{gg} \right),
\end{align}
Here $n_{1/2}$ denote the number of singly/doubly occupied sites determined by the initial density profile, respectively. A similar equation holds for $N_e$, and the experiment records $\eta_{\rm rp} N_e$. 

The outer shell of singly-occupied sites thus appears as a ``parasitic'' background signal on top of the atom pair signal of interest. This signal is however easy to correct for, since it merely corresponds to an ensemble of $n_1$ indenpendent two-level systems undergoing a Landau-Zener process. Provided one is able to determine $n_{1/2}$, the spurious contribution of singly-occupied sites can be substracted off the measurements to obtain the contribution of atom pairs. 

To measure $n_{1/2}$, we hold the atoms for a variable time $t_{\rm hold}$ in presence of laser light after the preparation ramp, and record the remaining $g$ population as a function of $t_{\rm hold}$. An example of such measurement is shown in Fig.\,\ref{fig2}. We fit such curves by an exponentially decaying function $f_1+f_2e^{-\gamma t_{\mathrm{hold}}}$ where $f_{1/2}=n_{1/2}$ in Eq.\,(\ref{eq:Ng}), and where $\gamma$ gives the decay rate of the prepared non-Hermitian dressed state.

\section{Strongly enhanced lifetime in the quantum Zeno regime}
\label{sec:lifetime}
\begin{figure}[ht!!!!!!]
\centering
\includegraphics[width=\columnwidth]{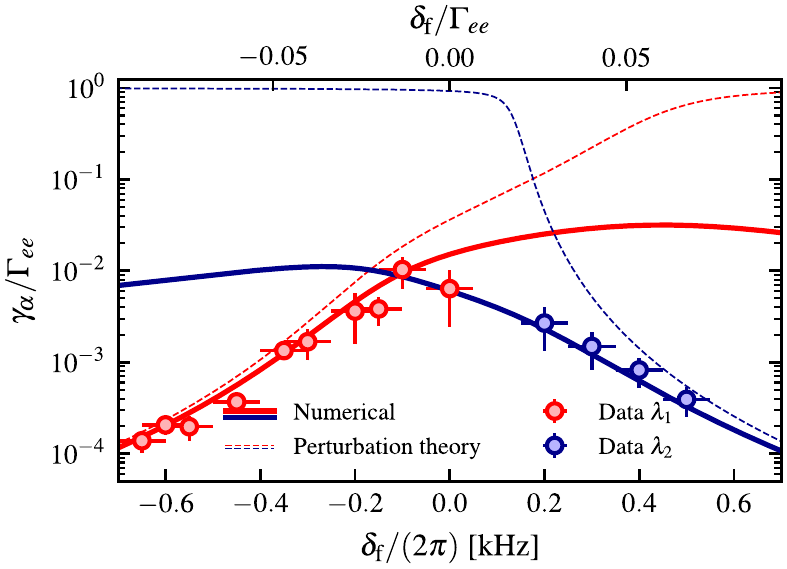}
\caption{Measurement of decay rates of the non-Hermitian dressed eigenstates. Solid: numerically calculated decay rates. Dashed: lifetime expected when treating the non-Hermitian part as a perturbation. }
\label{fig3}
\end{figure}

We first discuss the observed lifetime of the state obtained at the end of the Landau-Zener ramps. We postpone the discussion of (quasi-)adiabatic following for a non-Hermitian system to the following Section\,\ref{sec:adia}. We perform the preparation sequence described above \textit{(a)} starting from $\delta_{\rm i}/(2\pi)=-1.5\,$kHz and increasing the laser detuning, or \textit{(b)} starting from $\delta_{\rm i}/(2\pi)=+1.5\,$kHz and decreasing the laser detuning. In case \textit{(a)}, the initial state almost coincide with the dressed state $\lambda_1$ and the experimental ramp aims at following ``adiabatically'' this state. Case \textit{(b)} is the same but following $\lambda_2$. 

After the preparation ramp, we hold the atoms for a variable time $t_{\rm hold}$ in presence of laser light, and finally record the remaining $e$ population. For both ascending \textit{(a)} and descending \textit{(b)} ramps, monitoring the atom number as a function of hold time $t_{\rm hold}$ allows us to extract the lifetime of the prepared dressed state as discussed in Section\,\ref{sec:detection}. 

We report in Fig.\,\ref{fig3} the result of all lifetime measurements using the ascending and descending frequency ramps \textit{(a)} and \textit{(b)} for several final laser detunings.  We compare the measured decay rates to the calculated decay rates of the eigenstates $\lambda_1$ and $\lambda_2$. The agreement between the predictions and the measurements is excellent. This validates \textit{a posteriori} the description by a non-Hermitian Hamiltonian and supports the claim of adiabatic following (see also Section\,\ref{sec:adia} below). 

To highlight the role of the quantum Zeno effect, we also show as dashed lines the predictions of a naive perturbation theory, where the non-Hermitian dressed eigenstates are replaced by the corresponding eigenstates of $\hat{H}_0 \, + \, \hat{W}$ (\textit{i.e.} setting $\Gamma_{ee} = 0$ in the non-Hermitian Hamiltonian), and Eq.\,(\ref{eq:gamma_i}) is then used to compute the decay rates. This comparison shows that the lifetime is enhanced by two orders of magnitude near resonance by the quantum Zeno effect.

\section{Quasi-adiabatic ramps}
\label{sec:adia}

In this Section, we examine the notion of quasi-adiabaticity and relate it to our experiments. We first remind how the concept of adiabatic following in Hermitian quantum mechanics can be generalized for non-Hermitian systems\,\cite{nenciu92,Uzdin11,Ibanez14,milburn15,nasari2022a,Melanathuru2022a,singhal2022a}
, and thereby define the notion of (quasi-)adiabatic following in a dissipative system. In a second part, we explore in more details the quasi-adiabaticity of the experimental ramps discussed in the previous Section.

\subsection{Quasi-adiabaticity for non-Hermitian systems}

We now consider a time-dependent Hamiltonian $\hat{H}_{\mathrm{eff}}(t)=\hat{H}_{\mathrm{eff}}[\delta(t)]$. The time-dependence originates from the detuning $\delta(t)$ taken to be a linear function of time with slope $\dot{\delta}$. 
We note $\vert \lambda_n\rangle$ and $\langle \bar{\lambda}_n \vert$ the right and left eigenstates of $\hat{H}_{\mathrm{eff}}$ associated with the eigenvalue  $\lambda_n=\varepsilon_n-i\hbar\gamma_n/2$ ($ \gamma_n \geq 0$). We take the normalization conventions $\langle \bar{\lambda}_m \vert  \lambda_n \rangle =\delta_{mn}$,  $\langle \lambda_n \vert  \lambda_n \rangle =1$\,\footnote{For non-Hermitian systems, there are two arbitrary normalization factors that can be chosen freely\,\cite{brody2013a,singhal2022a}}. Assuming that the system is  initially prepared in one particular eigenstate $\alpha$ of $\hat{H}_{\mathrm{eff}}(0)$, the non-Hermitian generalization of adiabatic mapping is\,\cite{nenciu92,Uzdin11,Ibanez14,milburn15,nasari2022a,Melanathuru2022a,singhal2022a}
\begin{align}\label{eq:PsiadiaNH}
 \vert \Psi(0) \rangle = \left\vert \lambda_\alpha[\delta_\ii] \right\rangle \to   \vert \Psi(T_R) \rangle =\ee^{\ii\phi_\alpha-\frac{1}{2}\kappa_\alpha} \, \vert \lambda_\alpha[\delta_\ff] \rangle,
 \end{align}
with $T_R$ the ramp duration. The system follows the state $\left\vert \lambda_\alpha \right\rangle$ up to a phase, as in the Hermitian case, but also up to an attenuation factor $\ee^{-\kappa_\alpha}$. One may speak of ``quasi-adiabaticity'' if the time evolution is well described by Eq.\,(\ref{eq:PsiadiaNH}). By this, we mean that the evolution is as close as possible to a truly adiabatic one, given the dissipative nature of the system at hand. 

It is convenient to parametrize the instantaneous detuning as $\delta(x) = (1-x) \delta_{\ii} + x\delta_{\ff}$, or equivalently $x= \dot{\delta} t/(\delta_{\ff}-\delta_{\ii}) \in [0,1]$. The quasi-adiabatic phase and attenuation exponent can then be written as
 \begin{align}
\phi_\alpha & =\int_{0}^{1}   \left( -\frac{\varepsilon_\alpha[x] T_R}{\hbar} + \mathrm{Re}\mathcal{B}_{\alpha\alpha}[x] \right) dx ,\\
\kappa_\alpha & = \int_{0}^{1} \Big( \gamma_\alpha[x]T_R+ 2\mathrm{Im}\mathcal{B}_{\alpha\alpha}[x] \Big) dx,
 \end{align}
with $\mathcal{B}_{\alpha\beta}[x]= \ii \Big\langle \bar{\lambda}_\beta \Big\vert \frac{d\lambda_\alpha}{dx} \Big\rangle_x$ a Berry connection associated with ``transport'' of the eigenstates in the $\Omega-\delta$ space. Note that the explicit form of $\phi_\alpha$ and $\kappa_\alpha$ depend on the choice of the normalization (see Ref.\,\cite{singhal2022a} for a detailed discussion).

Since the norm of the wavefunction is not conserved by non-Hermitian evolution, the non-Hermitian dressed eigenstate is transported  with a \emph{survival probability} less than one,
\begin{align}\label{eq:Ps}
P_{\mathrm{s},\alpha}=\ee^{-\kappa_\alpha}.
\end{align}
Eq.\,(\ref{eq:Ps}) provides a lower bound  for the actual survival probability, which is realized only for quasi-adiabatic evolutions following the least dissipative eigenstate.

A generic validity criterion for the quasi-adiabatic mapping (\ref{eq:PsiadiaNH}) is given by the inequality\,\cite{Ibanez14}
\begin{align}\label{eq:adiaconditionNH}
\frac{1}{\hbar} \left\vert \left( \frac{d \hat{H}_{\mathrm{eff}}}{dt} \right)_{\beta \alpha} \right\vert = \frac{\vert\dot{\delta} \vert}{2} \left\vert \langle \bar{\lambda}_\beta \vert \hat{S}_z \vert \lambda_\alpha \rangle \right\vert \ll  
 \frac{ \vert \lambda_\alpha-\lambda_\beta \vert^2}{\hbar^2}  \frac{P_{\mathrm{s},\alpha}}{P_{\mathrm{s},\beta}} \,\,\, 
\end{align}
$\forall \beta \neq\alpha$. There are two main differences from the analogous Hermitian adiabatic condition. First, the equality of the \emph{real parts} $\epsilon_i$ is only a necessary condition to follow the initial dressed state. Second, the right-hand-side of Eq.\,(\ref{eq:adiaconditionNH}) has an extra factor $P_{\mathrm{s},\alpha}/P_{\mathrm{s},\beta}$. If $\alpha$ is the least dissipative eigenstate, this factor is positive and exponentially large for slow ramps: quasi-adiabaticity is then reinforced by the non-Hermitian dynamics\,\cite{nenciu92}. 

\subsection{Experimental ramps}

\begin{figure}[ht!!!!!]
\centering
\includegraphics[width=\columnwidth]{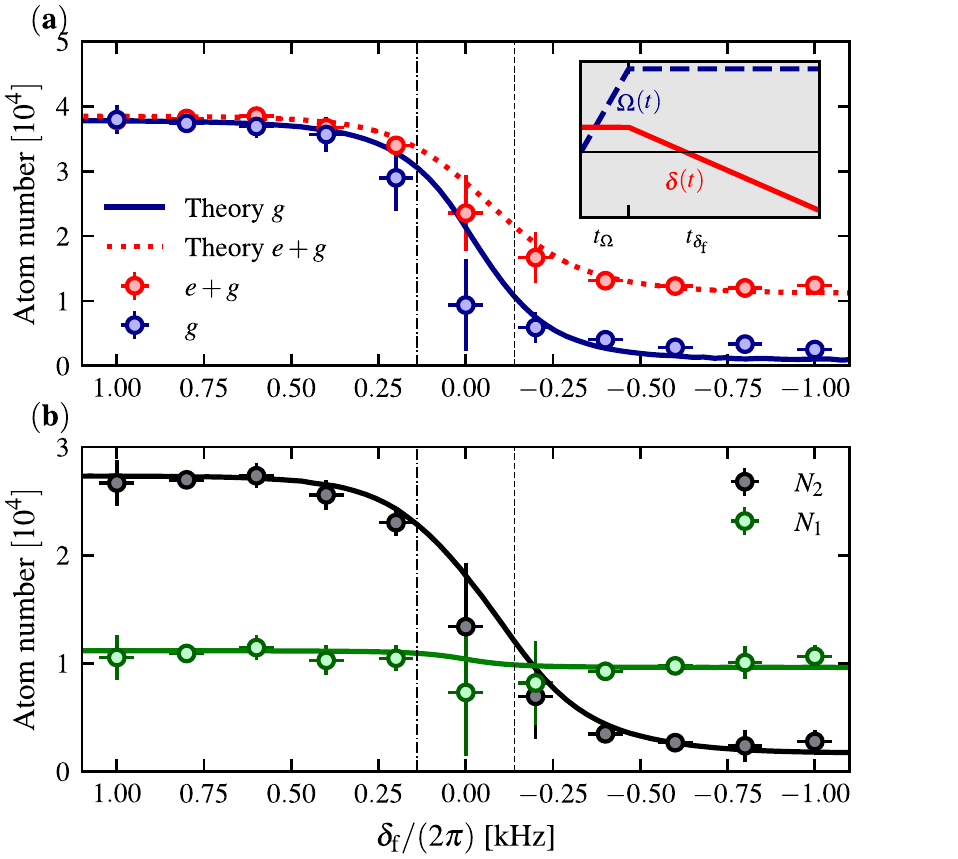}
\caption{(a): Total atom number and $g$ population following the $\lambda_2$ dressed state (descending frequency ramps at constant speed $\dot{\delta}$). Note that the frequency axis is reversed, so that the ramp time increases from left to right. 
The insets show a sketch of the ``trajectories'' in the $\Omega-\delta$ plane.  (b): Total populations $N_2$ and $N_1$ of the central core and of the outer shell, respectively. The former can be interpreted as a measure of the survival probability of doubly-occupied sites. We observe an excellent agreement with the expected population of the $\vert \lambda_2 \rangle$ dressed eigenstate (solid lines), which provides evidence for quasi-adiabatic following. The population $N_1$ of singly-occupied sites remains approximately constant. For such sites, the frequency ramp amounts to an adiabatic passage from $g$ to $e$ for large final detunings. }
\label{fig4}
\end{figure}

Fig.\,\ref{fig4}a shows the experimentally measured populations in $g$ and the total atom number for each final detuning $\delta_f$ of a \emph{descending} frequency ramp starting from a large and positive initial detuning $\delta_{\rm i}/(2\pi)=+1.5\,$kHz and lowering the detuning at constant speed $\dot{\delta}$. This ramp aims at following the $\vert \lambda_2 \rangle$ eigenstate of the non-Hermitian Hamiltonian, which connects to $\vert gg \rangle$ when $\delta \to + \infty$. We focus on descending ramps in this Section for concreteness, but we obtain equivalent results for ascending ramps following $\vert \lambda_1 \rangle$. The measured populations closely follow the curves expected from solving numerically the time-dependent Schroedinger equation for the experimental ramp.

The survival probability of doubly-occupied sites introduced previously provides a more direct experimental evidence for quasi-adiabatic following. From the decay measurements discussed in Section\,\ref{sec:detection}, we obtain the total populations $N_2=2n_2$ and $N_1=n_1$ of the central core and of the outer shell, respectively, for each value of the final detuning $\delta_\ff$. The resulting curves are shown in Fig.\,\ref{fig4}b. The former quantity can be interpreted as a measure of the survival probability of doubly-occupied sites times their initial total population. We observe that the measured $N_2$ agrees well  with the expected population of the $\vert \lambda_2 \rangle$ dressed eigenstate, \emph{i.e.} $N_2(\delta_\ff)= N_2(\delta_\ii) P_{s,\lambda_2}$. As discussed before, this  supports the interpretation of the experiment in terms of quasi-adiabatic following of the dressed eigenstate $\vert \lambda_2\rangle$. Note also that the population $N_1$ remains almost constant, as expected. The small drop in the calculated curve after crossing the resonance is due to the slightly lower detection efficiency for atoms in $e$, which is is included in the calculations.

Finally, we briefly mention that there are small deviations from adiabatic behavior in Fig.\,\ref{fig4}a, namely large fluctuations of the observables near the resonance, as well as a small residual $g$ population for large and negative detunings $\delta_\ff$. In the Supplementary Material\,\cite{supmat}, we argue that these deviations are caused by frequency fluctuations of the driving laser, and unrelated to the non-Hermitian dynamics of doubly-occupied sites that is the focus of this paper.

\section{Conclusion and discussion}

In conclusion, we have studied the non-Hermitian dynamics of atoms pairs tightly confined in an optical lattice. With strong two-body losses, we observe lifetimes much longer than the natural lifetime set by the inverse two-body loss rate $\Gamma_{ee}$ by at least two orders of magnitude. We discussed how this feature can be understood from the quantum Zeno effect, and demonstrated quasi-adiabatic preparation of the longest-lived eigenstate of the non-Hermitian Hamiltonian.

The non-Hermitian dressed state prepared after a quasi-adiabatic ramp is of the form
\begin{align}\label{eq:Psi_ent}
\vert \Psi \rangle & = \alpha \vert gg \rangle + \beta \vert ge \rangle.
\end{align}
 Paskauskas and You\,\cite{paskauskas2001a} have discussed criterions to decide whether a two-particle bosonic state shows quantum correlations, or, equivalently, is not separable. The state (\ref{eq:Psi_ent}) exhibits quantum correlations in the sense of Paskauskas and You unless either $\alpha$ or $\beta$ vanishes. 
 It is instructive to consider how the initially ``polarized'' (and separable) state $\vert g  g \rangle$ can be mapped to the state (\ref{eq:Psi_ent}) using Landau-Zener ramps. Let us for a moment neglect interactions and dissipation. The three-state system Hamiltonian $\hat{H}_0+\hat{W}$ then reduces to a spin-1 Hamiltonian involving only generators of SU$(2)$. Consequently, the associated evolution operator $\exp(-\ii \int_0^t \hat{H}(t')dt'/\hbar)$ is a rotation $\mathcal{R}$ transforming each spin$-1/2$ particle as $\vert g \rangle \to \vert \mathcal{R}g \rangle$. 
The final two-particle state retains a product form $\vert \mathcal{R}g\rangle \otimes \vert \mathcal{R}g \rangle$, and there are no quantum correlations\,\cite{paskauskas2001a}. The same conclusion holds for an interacting system with $U_{ee}+U_{gg}=2 U_{eg}$, \emph{i.e.} $q=0$ with our notations. 
 
For non-symmetric interactions leading to $q \neq 0$,  the evolution operator is no longer a mere rotation and can create correlations. For instance, if $U_{ee} \gg U_{eg},U_{gg}$, an adiabatic frequency ramp that stops well after crossing the $gg-eg$ resonance but well before crossing any other would generate a state of the form (\ref{eq:Psi_ent}).
Entanglement results in this situation from an interaction blockade phenomenon.

It is remarkable that a \emph{purely dissipative} evolution can accomplish the same thing. One may speak of ``dissipative blockade'' to highlight the analogy with the unitary evolution described above. The ideal case would correspond to an interaction-symmetric situation with $q=0$ but $\Gamma_{ee} \neq 0$, a situation close to the experimental one where $\hbar\Gamma_{ee} \gg \vert q \vert$. In that sense, one can consider the experiments reported here as a minimal instance demonstrating dissipation-assisted engineering of non-trivial quantum states, a topic of farther-reaching significance when applied to more complex many-body systems \cite{Muller12,Daley14,Diehl_2008,Verstraete_2009}. A related theoretical study on a related system of lossy fermions predicts, for instance, the possibility to create Dicke-like states \cite{fossfeig2012a}. An extension of the present work to a Hubbard-regime quantum gas, with non-negligible tunneling between neighboring wells, would bring further progress in this direction.  

To conclude this discussion, let us stress that the final state is actually an incoherent mixture of the two-particle entangled state (\ref{eq:Psi_ent}) and of the vacuum state (as in \cite{fossfeig2012a}). In the present work, we only perform ensemble measurements, and therefore only probe statistical mixtures. The entanglement could however be revealed in quantum gas microscope experiments\,\cite{ott2016a}, where the spin-resolved population of each site is accessible, and where post-selecting the doubly-occupied sites could make the entangled state experimentally detectable.  

%

\vskip 2mm
\begin{acknowledgments}
We thank Stefan Rotter and Leonardo Mazza for insightful discussions. LKB is a member of the network ``Science and Engineering  for Quantum Technologies in the Ile-de-France Region".
\end{acknowledgments}

\bibliography{bib_quasiadiabaticZeno_sub.bib}
\bibliographystyle{apsrev_nourl.bst}

\end{document}